\documentclass{nature}

\usepackage{graphicx}
\makeatletter
\let\saved@includegraphics\includegraphics
\AtBeginDocument{\let\includegraphics\saved@includegraphics}
\renewenvironment*{figure}{\@float{figure}}{\end@float}
\makeatother

\graphicspath{{./}{../Figures/}{../}}

\title{On stable H-C-N-O compounds at high pressure}

\author{Lewis J. Conway,$^{1}$ Chris J. Pickard,$^{2,3}$ \& Andreas Hermann$^1$}

\begin{document}

\maketitle

\begin{affiliations}
    \item Centre for Science at Extreme Conditions \& The School of Physics and Astronomy, The University of Edinburgh, Edinburgh EH9 3FD, United Kingdom 
    \item Department of Materials Science \& Metallurgy, University of Cambridge, Cambridge CB3 0FS, United Kingdom
    \item Advanced Institute for Materials Research, Tohoku University, Aoba, Sendai, 980-8577, Japan
\end{affiliations}

\begin{abstract}
The make-up of the outer planets, and many of their moons, are dominated by matter from the H-C-N-O chemical space, commonly assumed to originate from mixtures of hydrogen and the planetary ices H$_2$O, CH$_4$, and NH$_3$\cite{HUBBARD1991,Helled2011,Nettelmann2016,Helled2020}. In their interiors, these ices experience extreme pressure conditions, around 5~Mbar at the Neptune mantle-core boundary, and it is expected that they undergo phase transitions, decompose, and form entirely new compounds\cite{Hermann2012,Pickard2013,Gao2010,Qian2016}. In turn, this determines planets' interior structure, thermal history, magnetic field generation, etc. Despite its importance, the H-C-N-O space has not been surveyed systematically. Asked simply: at high-pressure conditions, what compounds emerge within this space, and what governs their stability? Here, we report on results from an unbiased crystal structure search amongst H-C-N-O compounds at 5~Mbar to answer this question.
\end{abstract}

Crystal structure prediction coupled to electronic structure calculations has emerged as a powerful tool in computational materials science, in particular in the area of high-pressure science, where it can overcome the chemical imagination attuned to ambient conditions: the predictions -- and subsequent experimental confirmations -- of unusual compounds such as Na$_2$He, H$_3$S, or LaH$_{10}$ attest to the predictive power of these approaches\cite{Dong2017,Li2014,Duan2015,Drozdov2015,Peng2017a,Geballe2018,Drozdov2019}. High-pressure phases of planetary ices have equally been explored computationally, and some predictions of exotic phases have also been confirmed by experiments\cite{Pickard2008,Gao2010,Hermann2012,Ninet2014,Palasyuk2014}.  However, with increasing number of constituents the structure searches become computationally much more demanding. Hence, while structure predictions for elemental and binary systems are routine, there are far fewer extensive searches of ternary systems\cite{Fredeman2011}, and none for quaternary systems.

The situation in the H-C-N-O chemical space reflects this. A vast number of publications exist on the high-pressure evolution of the individual constituents H, C, N, and O\cite{Pickard2007,Sun2012d,Martinez-Canales2012,Sun2013}. Binary phase spaces have also been looked at in great detail\cite{Pickard2013,Qian2016,Pickard2016,Conway2019}. The ternary phase spaces are much less investigated: while diagrams H-C-O, H-N-O and C-N-O phase diagrams have been reported\cite{Saleh2016,Shi2018,Steele2017}, the H-C-N space has not, for example. The PubChem database\cite{Kim2019} lists just under 4.8m molecular H-C-N compounds. Overall, it contains 44.6m H-C-N-O compounds, 78\% of which are of true quaternary composition -- this highlights both the complexity and the relevance of this composition space for organic chemistry. Back in the high-pressure area, some binary mixtures of planetary ices, for instance H$_2$O-NH$_3$ or N$_2$-CH$_4$ mixtures, which form a subset of the H-N-O and H-C-N spaces, have also been studied into the Mbar pressure range\cite{Bethkenhagen2015,NadenRobinson2017,NadenRobinson2018,Peng2020}. However, despite the overall importance of H-C-N-O both to planetary science and Earth-bound chemical and life sciences, to our knowledge there are no computational high-pressure studies of binary or ternary molecular systems that include all four elements -- for example, the CO$_2$-NH$_3$ or NH$_3$-H$_2$O-CH$_4$ systems -- let alone the full H-C-N-O system. Here, we explore the full H-C-N-O chemical space via crystal structure searches performed at 500~GPa, resembling the pressure at Neptune's core-mantle boundary, and therefore representative for the deep interior pressure conditions for the outer planets and giant icy exoplanets.

\section*{Results and Discussion}

\begin{figure}
	\includegraphics[width=\textwidth]{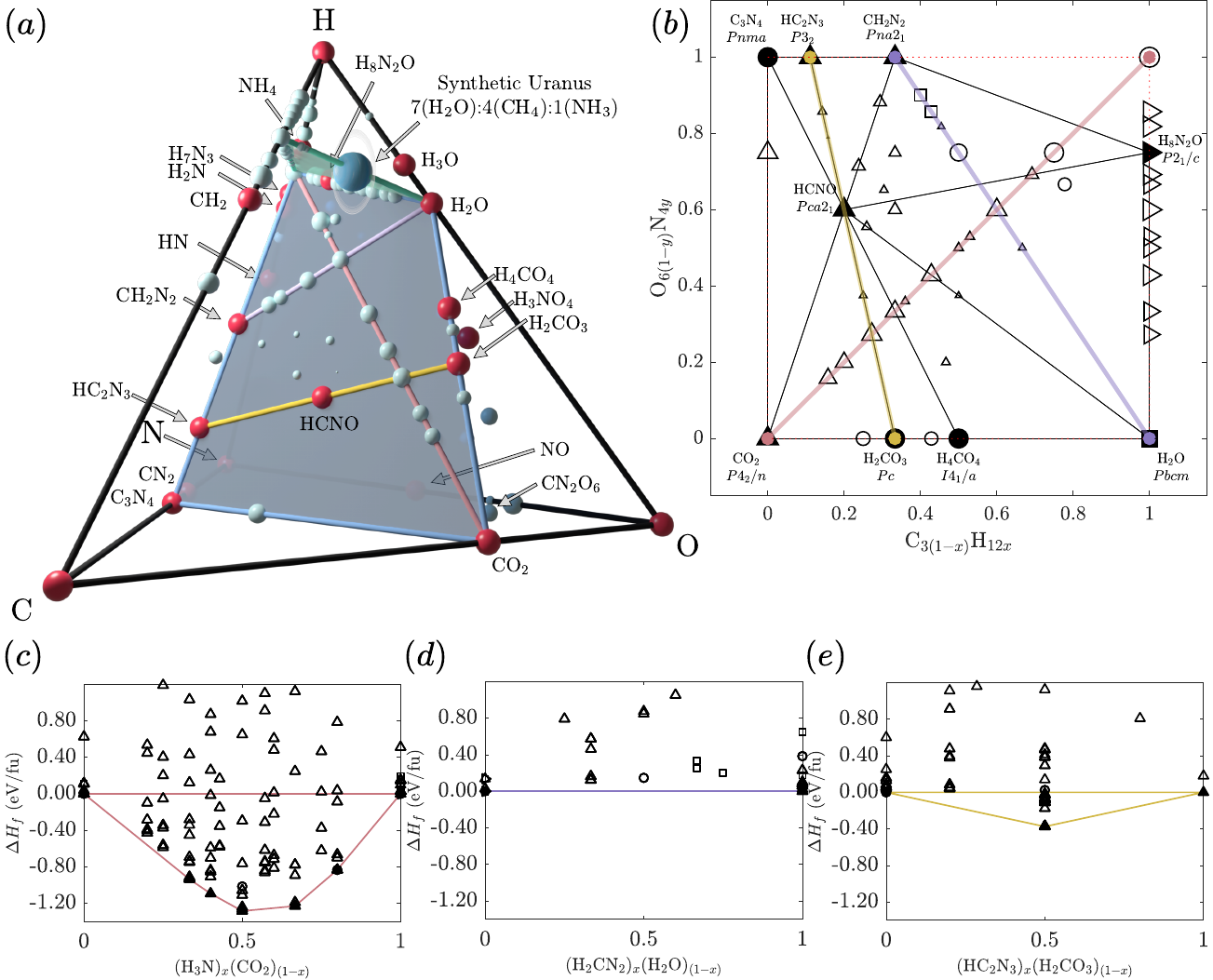}
 	\caption{Calculated H-C-N-O phase diagram at 500~GPa. (a) Full quaternary phase diagram, with the CO$_2$-C$_3$N$_4$-H$_2$O-NH$_3$ subspace shaded in grey, and the 'planetary ice' triangle H$_2$O-CH$_4$-NH$_3$ in green. Red/blue symbols are stable/metastable phases as labelled; for the latter, size represents closeness to stability. For reference, the 'synthetic Uranus' composition is marked. (b) The balanced redox CO$_2$-C$_3$N$_4$-H$_2$O-NH$_3$ subspace of H-C-N-O, with select internal 1D cross-sections highlighted. Full/open symbols are stable/metastable phases. Circles/upward triangles/sideways triangles are from searches of the full H-C-N-O space/the CO$_2$-C$_3$N$_4$-H$_2$O-NH$_3$ plane/the H$_2$O-CH$_4$-NH$_3$ plane; square symbols are manually added known structures. (c)-(e) Binary convex hulls from select 1D paths traversing the CO$_2$-C$_3$N$_4$-H$_2$O-NH$_3$ plane as shown in (b); enthalpies are relative to the binary end members which may not be stable in the full quaternary space. (c) CO$_2$-NH$_3$ phases; (d) CH$_2$N$_2$-H$_2$O phases; (e) HC$_2$N$_3$-H$_2$CO$_3$ phases.}
 	\label{fig:HCNO}
\end{figure}

The quaternary H-C-N-O phase diagram resulting from our searches is shown in Figure~\ref{fig:HCNO}(a), where all possible compounds are represented within (or on the surface of) a tetrahedron bounded by H, C, N, and O. We find only one truly quaternary compound to be stable, the 1:1:1:1 compound HCNO, at ambient conditions known as cyanic or fulminic acid, which we discuss in detail further below. We also find five new stable structures on three of the ternary faces: CH$_2$N$_2$ (cyanamide), H$_3$NO$_4$ (orthonitric acid), H$_8$N$_2$O (ammonia hemihydrate), HC$_2$N$_3$ (carbon nitride imide), and CN$_2$O$_6$.

A notable number of stable and metastable phases is located on a plane in the 3D composition space that is highlighted in Figure~\ref{fig:HCNO}(a) and bounded by CO$_2$, C$_3$N$_4$, H$_2$O and NH$_3$. This plane is drawn up in Figure~\ref{fig:HCNO}(b) along the axes CO$_2$--H$_2$O and CO$_2$--C$_3$N$_4$. These axes correspond to the conservation of valence electrons along the 'transmutation' C $\rightarrow$ 4H and the conservation of valence holes along the 'transmutation' 6O $\rightarrow$ 4N, respectively. The upper right corner of this plane is then ammonia, NH$_3$. As a consequence of conserving both electron and hole count, all compounds in this plane are the product of balanced reduction-oxidation (redox) reactions between `reducers' C and H and `oxidizers' O and N. Those roles are in line with the electronegativities of these elements at ambient conditions, as well as their estimates at high pressure.\cite{Rahm2019} Moreover, all \emph{stable} compounds found in this balanced redox plane fulfill the octet rule, where all constituents have filled outer electronic shells. Therefore, a first result from this study is that, even at 500~GPa, most of the stable compounds in the H-C-N-O chemical space (including HCNO) adhere to some of the classic stability criterions for chemical compounds. Furthermore, most relevant stable or metastable phases within this chemical subspace follow along simple binary mixture lines, which are shown in Figure~\ref{fig:HCNO}(c-e). A straightforward route to create quaternary high-pressure compounds would be from CO$_2$--NH$_3$ mixtures. We find five 'stable' mixtures along the binary phase space as shown in Figure~\ref{fig:HCNO}(c). However, these are all metastable against formation of HCNO; which in turn should be accessible as the 1:1 mixture of CO$_2$ and CH$_2$N$_2$, or HC$_2$N$_3$ and H$_2$CO$_3$. The situation at lower pressures is discussed further below.

\begin{figure}[ht]
	\includegraphics[width=6cm]{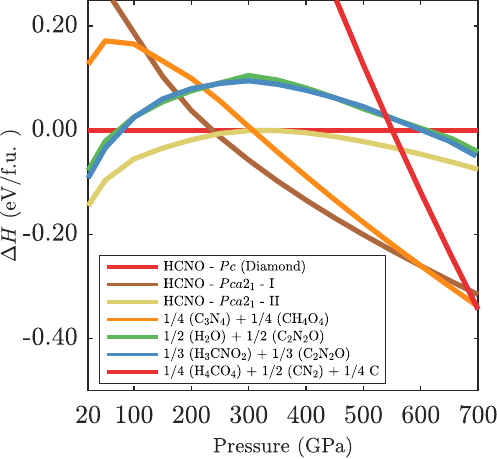}
    \includegraphics[width=5cm]{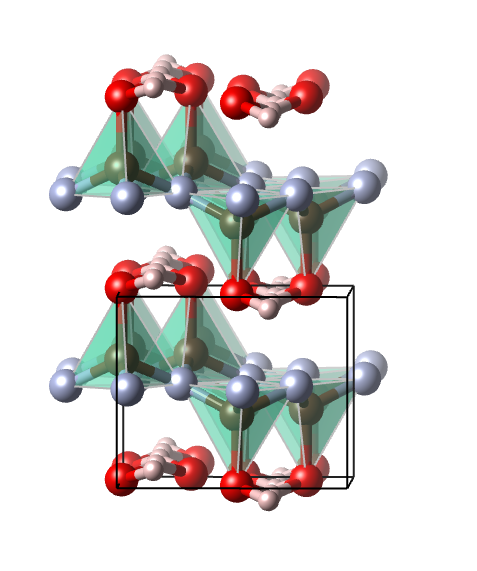}
	\includegraphics[width=4.5cm]{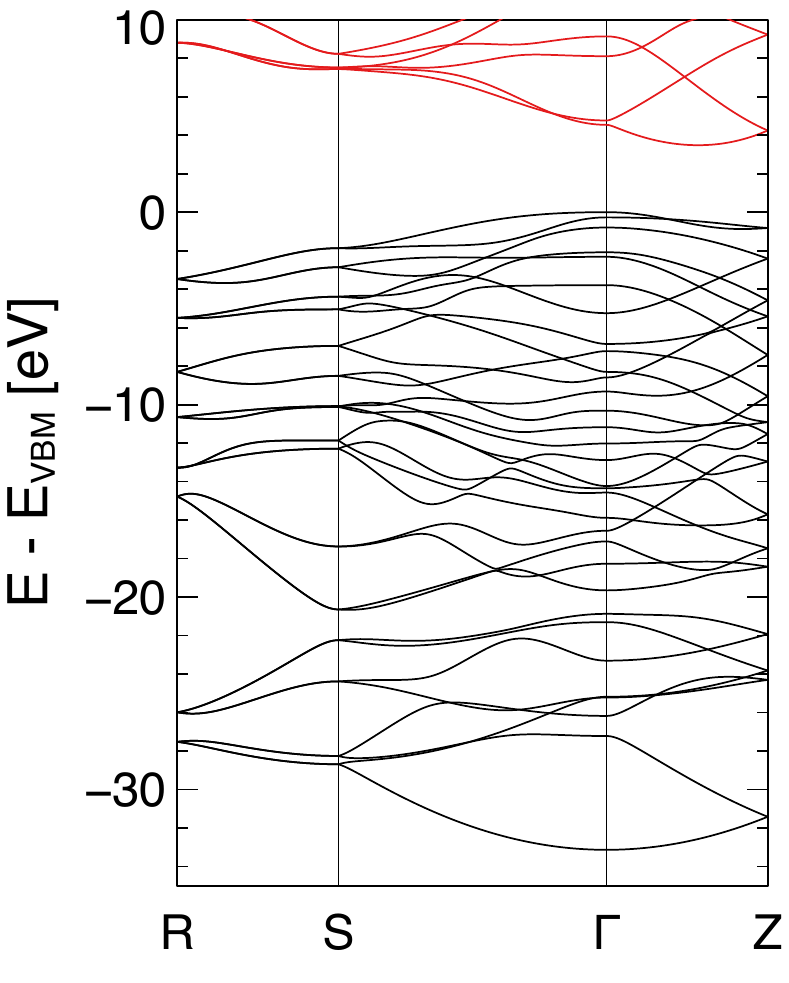}
	\caption{HCNO at 500~GPa. (a) HCNO enthalpy plot, including all relevant decomposition paths. (b) $Pca2_1$-II crystal structure, white/brown/blue/red spheres denote H/C/N/O atoms. Unit cell is shown, and tetrahedral CON$_3$ units are highlighted by polyhedra. (c) Electronic band structure of HCNO-$Pca2_1$-II.}
	\label{fig:HCNO-struc}
\end{figure}

The emergence of the simplest conceivable quaternary compound, HCNO, as a stable point in this chemical space might seem surprising. Fulminic/isocyanic acid, HNCO, was discovered in 1830 by Liebig and Wöhler\cite{Liebig1830}. It usually forms as H-N=C=O or as a tautomer, cyanic acid H-O-C$\equiv$N. These are also isomers of fulminic acid, H-C=N-O, which is an explosive. However, HCNO is isoelectronic to diamond, or carbon in general, and in the condensed state might form compact polyhedral networks that lead to stability under pressure. The stable orthorhombic high-pressure structure (denoted $Pca2_1$-II) is shown in Figure~\ref{fig:HCNO-struc} and appears to be closer to cyanic acid, with buckled graphitic C-N layers connected by C-O bonds to O-H chains. The layers are very close, and carbon is at the centre of tetrahedral CON$_3$ units. This structure is stable against decomposition between 240 and 600~GPa; at the low pressure end, it emerges from the HCNO-$Pca2_1$-I structure (see below and also the Supplementary Information, SI) and at the upper pressure end decomposes into 1/4(C$_3$N$_4$ + CH$_4$O$_4$). Partial atomic charges based on Bader's topological analysis of the electron density\cite{AIM.Bader1994} are 0.6, 2.2, -1.6 and -1.2~e for H, C, N and O respectively.  One can roughly identify these with formal oxidation states +1/+4/-3/-2, and conclude that ionic bonding contributes to HCNO's stability. HCNO is an insulator, with an indirect band gap of 3.73~eV at 500~GPa, see Figure~\ref{fig:HCNO-struc}(c). 

HCNO is not only isoelectronic to diamond, but also has low-energy polymorphs that are isostructural with diamond. A monoclinic $Pc$ phase with a cubic diamond topology is very close to stability around 250~GPa, see Figure~\ref{fig:HCNO-struc}(a). In fact, this structure can be seen as a quaternary variant of zinc blende (or cubic boron nitride), with alternating sublattices occupied by cations C and H, and anions O and N, respectively. Within the H-C-N-O space, the (HC)$_m$(NO)$_m$ subspace contains the compositions that would in theory allow the formation of such quaternary zinc blende structures. The intersection of this subspace with the balanced redox plane is highlighted by the yellow line in Figures~\ref{fig:HCNO}(a) and (b), which hosts three stable structures: besides HCNO, these are H$_2$CO$_3$, carbonic acid, and HC$_2$N$_3$, carbon nitride imide.

\begin{figure}[th]\centering
	\includegraphics[width=0.95\columnwidth]{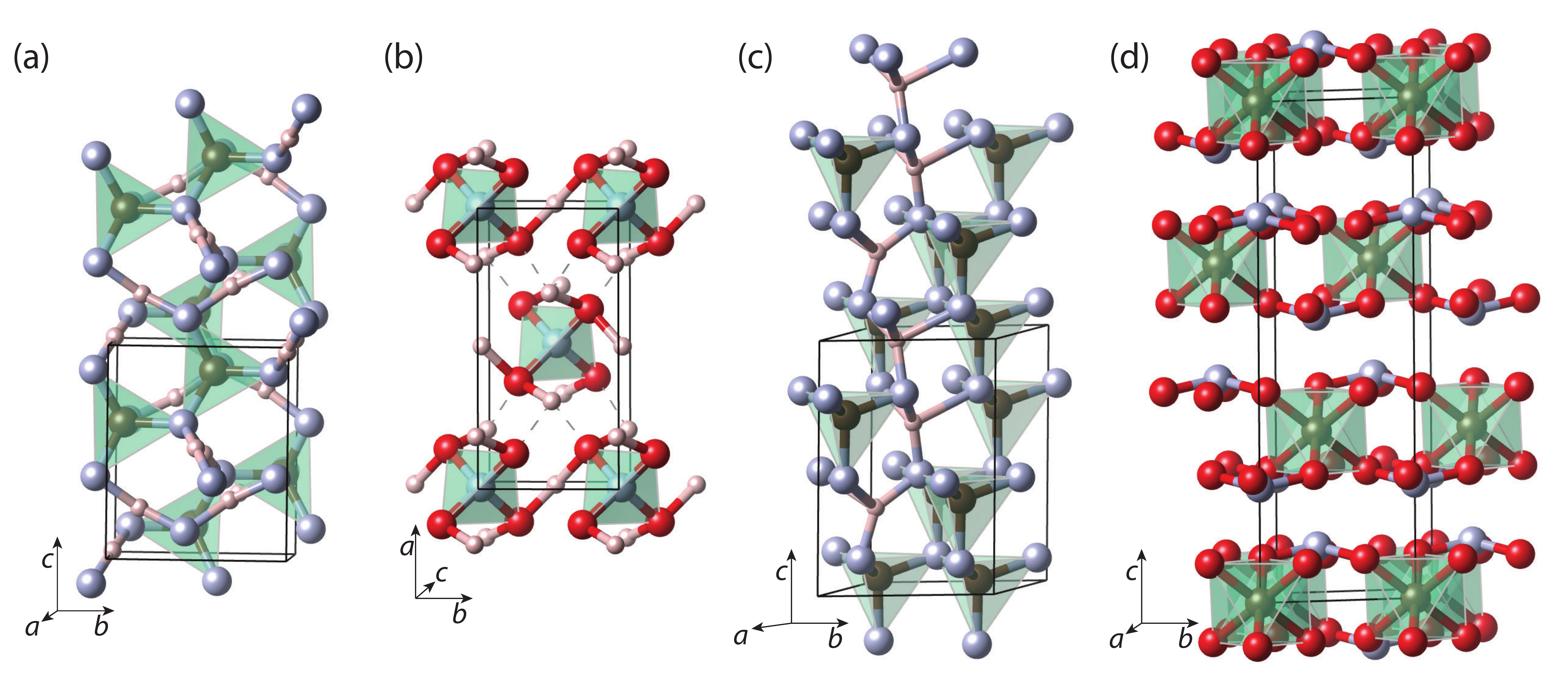}
	\caption{Crystal structures of ternary H-C-N-O phases, with carbon or nitrogen coordination highlighted. (a) CH$_2$N$_2$ (with CN$_4$ tetrahedra). (b) H$_3$NO$_4$ (with NO$_4$ tetrahedra). (c) HC$_2$N$_3$ (with CN$_4$ tetrahedra). (d) CN$_2$O$_6$ (with CO$_6$ octahedra).}
	\label{fig:ternary-struc}
\end{figure}

In total, five new stable ternary compounds emerge from the searches at high pressures. The structures of new stable stoichiometries are shown in Figure~\ref{fig:ternary-struc} (ammonia hemihydrate is discussed in the SI). Cyanamide, CH$_2$N$_2$, forms a molecular crystal at ambient pressure conditions.\cite{Denner1988a} It can exist as two tautomers: either as the planar molecule N$\equiv$C-NH$_2$ or as H-N=C=N-H, called carbodiimide. The high pressure structures for CH$_2$N$_2$ found here become stable above 10~GPa, see the SI for enthalpy data, and are networks structures dominated by C-N covalent bonds and NH$\cdots$N hydrogen bonds. As pressure increases, the N--H$\cdots$N hydrogen bond length decreases and symmetric bridging N--H--N bonds form (see Figure~\ref{fig:ternary-struc}(a)). The partial charges on C/N/H of 2.0/-1.6/0.5~e obtained from a Bader analysis are consistent with formal oxidation states +4/-3/+1 and cyanamide remains an insulator with a band gap of 3.2~eV at 500~GPa. Above 600~GPa CH$_2$N$_2$ transforms into a $C2/c$ phase that is isostructural with cubic diamond (see the SI), with only the H positions deviating noticeably from tetrahedral sites.

Another stable compound is H$_3$NO$_4$. At ambient pressures it can be seen as a nitric acid-water complex, HNO$_3\cdot$H$_2$O, which is present in polar stratospheric clouds, contributes to ozone depletion\cite{Canagaratna1998} and forms an ionic structure (NO$_3$)$^-\cdot$(H$_3$O)$^+$ at low temperatures\cite{Delaplane1975}. This compound is energetically stable at pressures above 200~GPa (see the SI). The structure consists of NO$_4$ tetrahedra in a body-centered tetragonal arrangement. The terminal oxygen atoms are connected by symmetric buckled O-H-O bonds to form an overall layered structure. The partial charges on the N/O/H atoms are 1.0/-0.8/0.7~e, thus overall slightly less ionic in character than CH$_2$N$_2$, in line with relatively close electronegativities of N and O. However, this compound is electronically very stable, with an unusually large band gap of 5.7~eV at 500~GPa. 

A third new compound HC$_2$N$_3$ is stable accross the pressure range in a $P3_2$ structure similar to hexagonal diamond. This is perhaps unsurprising as it has a relatively high carbon content and is isoelectronic to diamond. In fact, this compound has been reported in high-pressure/high-temperature syntheses in a `defective wurtzite' structure.\cite{Horvath-Bordon2007,Salamat2009} The $P3_2$ structure is more stable than the reported $Cmc2_1$ structure above 209 GPa. The partial charges on C/N/H are 2.0/-1.5/0.6~e, very similar to CH$_2$N$_2$, and the compound has a band gap of 4.0~eV at 500~GPa.

A slightly more unusual hydrogen-free CN$_2$O$_6$ is predicted stable above 600~GPa (see the SI). It is the only new stable compound not in the balanced redox subspace.  Its structure has a trigonal unit cell dominated by CO$_6$ octahedra connected by bridging nitrogen atoms such that they form layers in the {\em ab} plane.  Notably, this compound exhibits long sought-after octahedral coordination of carbon, at significantly lower pressures than in CO$_2$, where it is predicted to occur at 1000~GPa\cite{Lu2013}. The partial charges of the C/N/O atoms are 2.2/0.9/-0.7~e but ionicity is only part of the story, as this is a metallic compound. CN$_2$O$_6$ is two electrons short of a filled electronic shell; we constructed quaternary compounds (Be,Mg)CN$_2$O$_6$ by placing Be/Mg atoms between the layers of CN$_2$O$_6$. These compounds are stable against decomposition and insulating with band gaps of 4.91 and 2.91~eV respectively (see the SI).

\begin{figure}[th]\centering
	\includegraphics[width=0.99\columnwidth]{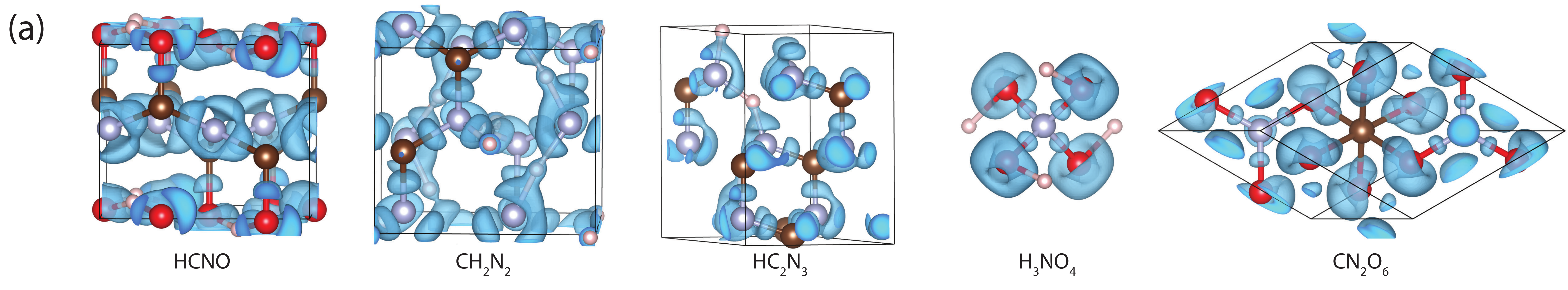}
	\includegraphics[width=0.99\columnwidth]{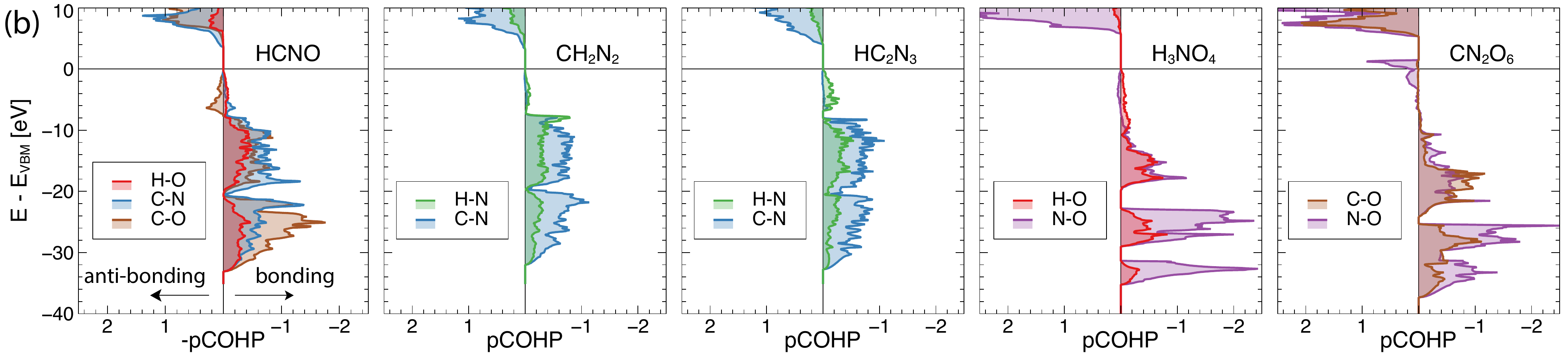}
	\caption{Chemical bonding analyses. (a) Electron localization function isosurfaces (ELF = 0.80) for stable H-C-N-O phases at 500~GPa, drawn to the same scale. (b) Crystal Orbital Hamilton Population (COHP) analysis for the same structures, projected onto different bond types as indicated.}
	\label{fig:elf-cohp}
\end{figure}

The new stable ternary or quaternary H-C-N-O stoichiometries are all relatively hydrogen-poor. Orthocarbonic acid, H$_4$CO$_4$,\cite{Saleh2016} and ammonia hemihydrate, H$_8$N$_2$O,\cite{NadenRobinson2017} stand out as the most hydrogen-rich ternary phases with 44 and 73 at-\% (5 and 15 wt-\%) hydrogen, respectively. Any new ternary or quaternary phases need to compete with obvious thermodynamic sinks such as carbon dioxide, water or polyethylene (CH$_2$), which all benefit from strong covalent and/or ionic bonding. Hence, a second general result from this study is that at 5~Mbar stable (new) structures tend to form covalently bonded polyhedral networks; they require a significant amount of the heavy elements to achieve this: C and N as polyhedra formers, and O (or N) as terminal atoms; their hydrogen content is then relatively low. These network structures are very compact, which favours their formation under pressure, and they tend to feature significant partial charge transfer, resulting in ionic bonding and (with exception of CN$_2$O$_6$) insulating character. Figure~\ref{fig:elf-cohp} shows the chemical bonding as analysed in real space (via the electron localization function, ELF)\cite{Becke1990,Savin1992} and in reciprocal space (via the Crystal Orbital Hamilton Population, COHP)\cite{Dronskowski1993,Maintz2013}. The ELF shows that all stable structures have strong covalent bonds between the heavy atoms, and filled electronic shells around O and N anions. The COHP corroborates that almost all interactions are bonding, with significant individual strength, including O-H and N-H bonds where hydrogen is present. Only CN$_2$O$_6$ and HCNO have some small antibonding character around the Fermi energy and valence band maximum, respectively. 

Mixtures of the planetary ices H$_2$O-CH$_4$-NH$_3$ form part of the H-C-N-O chemical space. However, the only stable structures in this `planetary ice triangle' at 500~GPa are H$_2$O and AHH -- CH$_4$ is unstable against decomposition into CH$_2$ and H$_2$, and NH$_3$ against NH$_4$ and N$_3$H$_7$. The initial searches did not produce anything -- regardless of stability --  on the interior of this triangle. Targeted searches of this plane revealed a previously unseen phase of ammonia hemihydrate (AHH). The structure is more stable than those found in previous searches\cite{NadenRobinson2018} and extends the maximum stability of AHH from 500~GPa to upwards of 800~GPa; see the SI for more details. The only stable compound with additional hydrogen is H$_3$O, where we confirm a recently reported structure.\cite{Huang2020} The predicted fate of complex icy mixtures depends on composition: a 1:1:1 mixture of H$_2$O:CH$_4$:NH$_3$ is predicted to decompose into H$_3$O:CH$_2$:NH$_4$, with $\Delta H_f = 0.33$~eV/molecule, while a 'synthetic Uranus' mixture of solar composition ratio 7:4:1 is slightly less unstable, with $\Delta H_f = 0.25$~eV/molecule against the preferred decomposition into 7*H$_3$O:4*CH$_2$:NH$_4$ (both at 500~GPa). Neither case involves the formation of free hydrogen (which could be balanced by formation of some of the stable compounds rich in heavy elements we find here). A third result from this extensive search is therefore that realistic molecular mixtures for icy planet interiors do not easily decompose into a hydrogen-rich and a heavy atom-rich component; the `diamond rain' predicted from the decomposition of methane into diamond and hydrogen\cite{Ross1981} is not favoured based on ground state energetics. In fact, seen as part of the full H-C-N-O space, this process might be much less relevant. The predicted formation of polyethylene, CH$_2$, from methane, releases hydrogen, but this would be absorbed into H$_3$O and NH$_4$, both of which are host-guest compounds where molecular H$_2$ is stored in H$_2$O and NH$_3$ host matrices, respectively\cite{Qian2016,Huang2020}. Dynamic compression experiments of polyethylene revealed it to be much more resistant to diamond formation at planetary conditions than, for example, polystyrene (CH)\cite{Hartley2019,Kraus2017}. On the other hand, quasi-harmonic free energy calculations showed that pressures for diamond formation from methane should be much lower at high temperatures\cite{Gao2010,Conway2019}. More studies are needed on whether this also holds for realistic molecular mixtures.

The SI gives further details on outcomes from the structure searches at 500~GPa for all ternary and binary subspaces, and detailed comparisons to literature data. In addition it discusses the H-C-N ternary space in more detail that has not yet been covered in the literature.

Our search was targeted at 500~GPa as a typical pressure deep inside Neptune-like planets. Less extreme conditions are also present, and also more easily accessible in experiments. A simple linear extrapolation of enthalpy changes, $\Delta H \approx \Delta p \cdot V_0$, works surprisingly well to identify competitive structures at pressures $p=p_0+\Delta p$, based purely on search results at $p+0=500$~GPa.\cite{Pickard2011} This method succeeded in finding lower pressure isomers of HCNO and CH$_2$N$_2$.  We then fully relaxed all potentially interesting structures over a wide pressure range to monitor structural changes, phase transitions, and establish lower bounds of stability. In addition we augmented the dataset by known compounds relevant at lower pressures. While our dataset might not contain all stable structures at lower pressures this analysis should still give an indication of how the H-C-N-O space evolves below 5~Mbar. For instance, the compound HCNO is found to be stable down to at least 20~GPa.

\begin{figure}[ht]\centering
    \includegraphics[width=7.5cm]{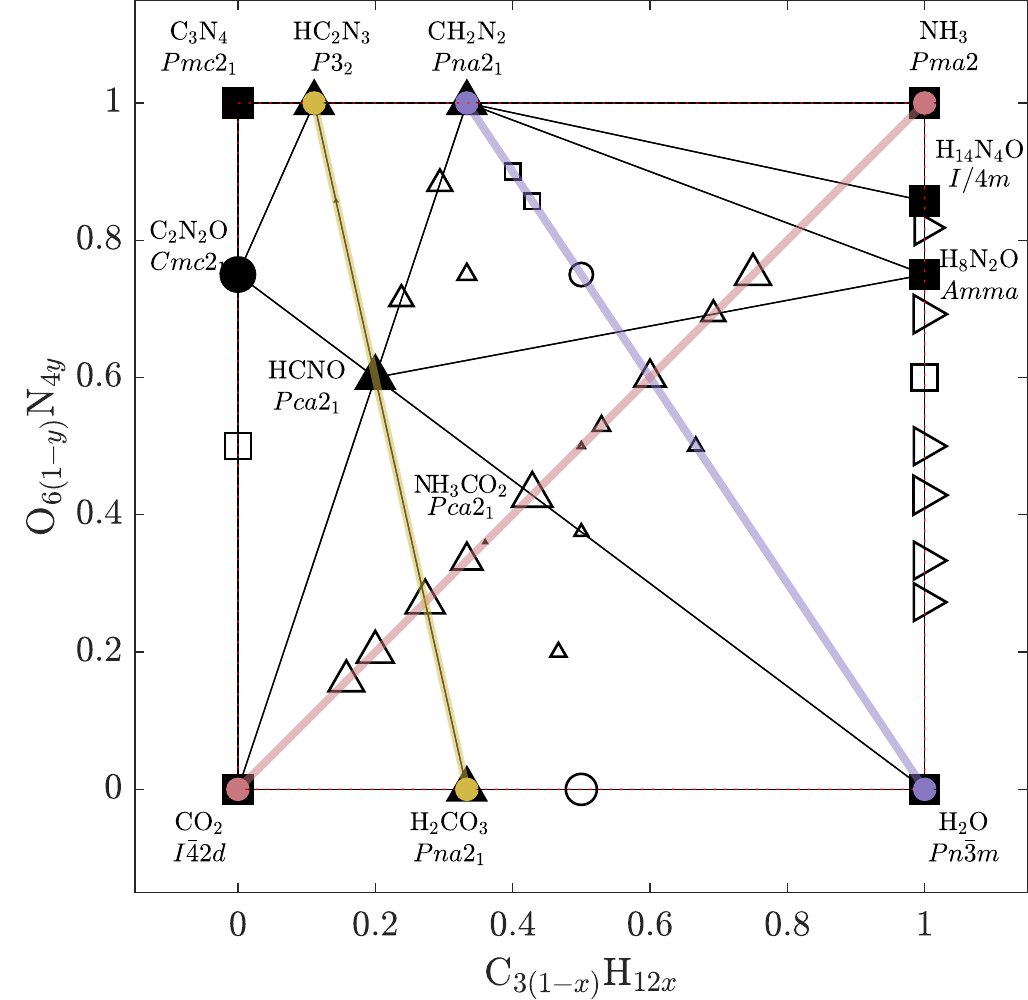}
    \includegraphics[width=7cm]{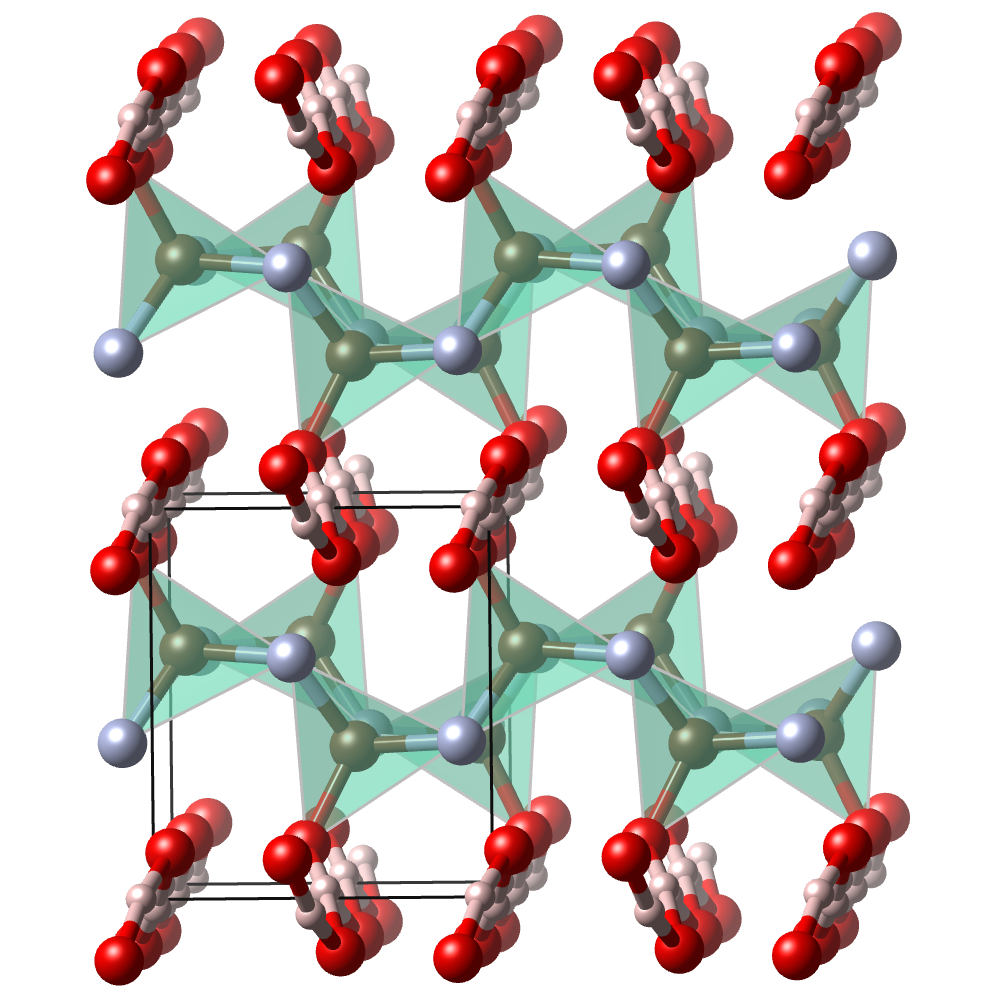}
	\caption{Calculated H-C-N-O phase diagram at 100~GPa. (a) Chemical subspace of `balanced redox'-compliant compounds, drawn to the same specifications as in Figure~\ref{fig:HCNO}(b).  (b) Crystal structure of HCNO-$Pca2_1$-I at 100~GPa, with unit cell and carbon coordination indicated.}
	\label{fig:100GPa}
\end{figure}

In Figure~\ref{fig:100GPa} we show the `balanced redox' plane spanned by CO$_2$, H$_2$O, NH$_3$, and C$_3$N$_4$ at 100~GPa. As at 500~GPa, this plane contains most stable or metastable ternary and quaternary phases. Along the edges, we find stable carbonic acid, H$_2$CO$_3$, ammonia hemihydrate (NH$_3$)$_2$(H$_2$O) and quarterhydrate (NH$_3$)$_4$(H$_2$O), cyanamide, CH$_2$N$_2$, carbon nitride imide, HC$_2$N$_3$, and oxycyanide, C$_2$N$_2$O, in agreement with literature\cite{Saleh2016,Steele2017,NadenRobinson2018}. All are simple binary mixtures of the end members of this chemical subspace, forming in 1:1, 1:2, or 1:4 ratios. The low pressure HCNO-$Pca2_1$-I structure remains stable at this pressure and down to at least 20~GPa (see SI), and remains the only stable truly quaternary compound. 

Note that the CO$_2$-NH$_3$ binary system (pink line in Figure~\ref{fig:100GPa}) features a total of seven `stable' phases at 100~GPa (see the SI for details). However they are metastable within the full H-C-N-O space because of the presence of HCNO. For instance, the 1:1 mixture, CH$_3$NO$_2$ known as nitromethane, is a widely used solvent and fuel additive at ambient conditions and lies just 10~meV/atom away from the hull at 100~GPa. Likewise, the 2:1 mixture, nitroacetic acid C$_2$H$_3$NO$_4$, is only 17~meV/atom above the hull. Finally, the 1:2 mixture CO$_2$ + 2NH$_3$, which forms ionic ammonium carbamate CO$_2$(NH$_2$)$^-\cdot$(NH$_4$)$^+$ at low temperatures, should decompose into three stable phases (HCNO, water, and ammonia hemihydrate) at 100~GPa, but can also be represented as 2H$_2$O + CH$_2$N$_2$ (purple line in Figure~\ref{fig:100GPa}). These CO$_2$-NH$_3$ mixtures (or the chemical compounds they form) can thus be used to explore the H-C-N-O space -- all of them should produce HCNO under pressure. Conversely, they could be used to probe the influence of kinetics {\emph vs} thermodynamics, as kinetic barriers can be significant in reactions of molecular compounds.

The structure of HCNO-$Pca2_1$-I is also shown in Figure~\ref{fig:100GPa}. It is similar to HCNO-$Pca2_1$-II at higher pressure, with stronger buckled graphitic C-N layers connected to --(O--H)-- chains along the $c$ direction. Partial charges on H/C/N/O of +0.66/+1.7/-1.15/-1.2 electrons are also very similar, suggesting similar chemical bonding, while a band gap of 5.5~eV at 100~GPa suggests strong electronic stability.

\section*{Conclusions}

In summary, we discuss here results from the first unbiased structure search of the H-C-N-O chemical space at high pressure, which dominates the interiors of icy planets. At pressure conditions that resemble those close to Neptune's core boundary, we find the simplest conceivable quaternary compound, HCNO, to be stable. In addition, we find several other stable ternary compounds, two of which (HC$_2$N$_3$ and CH$_2$N$_2$) are part of the hitherto unexplored H-C-N ternary space. From our analyses of the distribution of stable structures in chemical space and their chemical bonding we derive various conclusions.

Firstly, most relevant compounds adhere to the octet rule of filled electronic shells and balanced redox reactions, even at 500~GPa. We identify a two-dimensional subspace of H-C-N-O that is spanned by the transmutations C$\leftrightarrow$4H and 6O$\leftrightarrow$4N, which respectively conserve the number of valence electrons and valence holes, and which contains most relevant stable and metastable compounds. HCNO itself is part of this subspace. Secondly, stable compounds of three or four elements are relatively hydrogen poor. Such compounds can form compact polyhedral networks supported by strong covalent and ionic bonding - that seems to be necessary to retain stability against thermodynamic sinks such as water, carbon dioxide, or polyethylene. Thirdly, realistic mixtures of planetary ices do not easily decompose into these heavy atom-dominated compounds and pure hydrogen. Instead, based on ground state energetics, hydrogen released by methane forming polyethylene should be absorbed in host-guest networks of the type (H$_2$O)$_2$H$_2$ and (NH$_3$)$_2$H$_2$; the 'diamond rain' of heavy atom material falling through hydrogen-rich matter remains is not borne out by ground state energetics but remains a possible high-temperature effect.

At lower pressures, other compounds become relevant, and we suggest to use CO$_2$-NH$_3$ mixtures as springboard to explore the formation of truly quaternary H-C-N-O compounds in high-pressure environments. An obvious route to explore experimentally is the role of high temperatures: it can change relative free energies, can be used to investigate kinetic barriers towards the formation of the compounds predicted here, and can induce thermal excitations in these compounds  (superionicity, melting) that are relevant along typical icy planet isentropes. Looking beyond the H-C-N-O space, the addition of other constituents such as sulfur could be explored; the PubChem database lists 19.8m molecular compounds in the extended HCNO+S composition space. High pressure leads to different effects there: a C-S-H material was reported to superconduct at room temperature in a high-pressure phases\cite{Snider2020}, and metallicity and superconductivity is also seen in nitric sulfur hydride N-S-H mixtures\cite{Li2020c}.

\begin{methods}

All structure searches were carried out using the AIRSS and CALYPSO codes\cite{Pickard2011,Wang2012}, generating close to 400,000 structures in total. These were acquired in a sequence that initially searched the entire H-C-N-O space, followed by more targeted searches of several chemical subspaces, with around 280,000 structures in total. A further 100,000 structures were generated along the CH$_4$:NH$_3$ binary for the range of 4:1 to 1:4 mixtures at 100, 300 and 700~GPa. 

All geometry optimization and phonon calculations were carried out using the {\sc CASTEP} code\cite{Clark2005}, the PBE \cite{Perdew1996} exchange-correlation functional, and ultrasoft pseudopotentials generated on-the-fly by CASTEP. During the searches, geometry optimizations were performed with plane-wave wave cutoff of 340~eV and a k-point spacing of $0.7\times \pi$\AA$^{-1}$. After screening, a more precise calculation was performed with plane-wave wave cutoff of 1000~eV and a k-point spacing of $0.4\times \pi$\AA$^{-1}$.

Precise calculations were carried out on all structures on or close to the five dimensional convex hull of enthalpy against composition at 500~GPa. To estimate stable structures at lower and higher pressures, members of the two dimensional convex hull (for each composition) formed of enthalpy against volume were also calculated to a higher precision. Geometry optimizations were performed at pressures between 100 and 700~GPa in increments of 50~GPa. The enthalpies from these calculations were used to determine stable structures. All structures known from the literature which have not been reproduced in this search were manually added to the dataset. 

Phonon calculations were carried out on the stable structures to determine their dynamic stability using density functional perturbation theory (DFPT)\cite{Refson2006}, with norm-conserving pseudo potentials.

Bader charge, ELF, and COHP analyses were performed using wave functions obtained with the {\sc VASP} code in conjunction with the projector augmented wave (PAW) method\cite{VASP.Kresse1996,PAW.Joubert1999} and the PBE exchange-correlation functional\cite{Perdew1996}. Analyses were carried out using the {\sc CRITIC2} and {\sc LOBSTER} packages\cite{Otero-De-La-Roza2014,Nelson2020}.

\end{methods}




\begin{addendum}
 \item L.J.C. was supported by the UK’s EPSRC through the Condensed Matter Centre for Doctoral Training (EP/ L015110/1). C.J.P. was supported by the Royal Society through a Royal Society Wolfson Research Merit award. Computational resources provided by the UK’s National Supercomputer Service through the UK Car-Parrinello consortium (EP/P022561/1) and project ID d56 `Planetary Interiors’ and by the UK Materials and Molecular Modelling Hub (EP/P020194) are gratefully acknowledged.
 \item[Competing Interests] The authors declare that they have no competing financial interests.
 \item[Correspondence] Correspondence and requests for materials
should be addressed to A.H.~(email: a.hermann@ed.ac.uk).
\end{addendum}


\end{document}